\documentclass[twocolumn,preprintnumbers,showpacs,prb,amsmath,amssymb]{revtex4}

\usepackage{graphicx}
\usepackage{dcolumn}
\usepackage{bm}

\usepackage{hyperref}

\graphicspath{{Figures/}} 

\newcommand{\carb}{$^{13}\textrm{C}$ }

\begin{document}

\title{High resolution spectroscopy of single NV defects coupled\\
 with nearby $^{13}{\rm C}$ nuclear spins in diamond}
\author{A. Dr\'eau$^{1}$}
\author{J.-R. Maze$^{2}$}
\author{M. Lesik$^{3}$}
\author{J.-F.~Roch$^{3}$}
\author{V.~Jacques$^{1}$}
\email{vjacques@lpqm.ens-cachan.fr}
\affiliation{$^{1}$Laboratoire de Photonique Quantique et Mol\'eculaire, Ecole Normale Sup\'erieure de Cachan and CNRS UMR 8537, 94235 Cachan, France}
\affiliation{$^{2}$Facultad de F\'{i}sica, Pontificia Universidad Cat\'{o}lica de Chile, Santiago 7820436, Chile}
\affiliation{$^{3}$ Laboratoire Aim\'e Cotton, CNRS UPR 3321 and Universit\'e Paris-Sud, 91405 Orsay, France}

\begin{abstract}
We report a systematic study of the hyperfine interaction between the electron spin of a single nitrogen-vacancy (NV) defect in diamond and nearby \carb nuclear spins, by using pulsed electron spin resonance spectroscopy. We isolate a set of discrete values of the hyperfine coupling strength ranging from 14 MHz to 400 kHz and corresponding to \carb nuclear spins placed at different lattice sites of the diamond matrix. For each lattice site, the hyperfine interaction is further investigated through nuclear spin polarization measurements and by studying the magnetic field dependence of the hyperfine splitting. This work provides informations that are relevant for the development of nuclear-spin based quantum register in diamond.
\end{abstract}


\maketitle

\section{Introduction}
Owing to its outstanding optical and electron spin properties, the negatively charged nitrogen-vacancy (NV) color center in diamond has recently emerged as a promising candidate for a broad range of applications including quantum information processing (QIP),~\cite{Jelezko_PRL2004,Dutt_Science2007,Fuchs_NatPhys2011,Robledo_Nature2011,Neumann_Science2008} photonic devices,~\cite{Babinec_NatNano2010,Becher_NatNano2012,Faraon_NatPhot2011} hybrid quantum systems,~\cite{Japan_Nature2011,Kubo_PRL2011,Arcizet_NatPhys2011} nanoscale magnetometry,~\cite{Gopi_Nature2008,Maze_Nature2008,Maletinsky_2012,Rondin_2012} and imaging in life science.~\cite{Hollenberg_NatNano2011, Treussart_Small2011} For QIP applications, coherent interactions of the NV defect with nearby nuclear spins of the diamond lattice provide a fertile resource for the realization of universal quantum gates,\cite{Jelezko_PRL2004} quantum registers~\cite{Dutt_Science2007,Fuchs_NatPhys2011,Robledo_Nature2011} and multipartite entangled states.~\cite{Neumann_Science2008} For such experiments, individual nuclear spins are read out optically by coherently mapping their states onto the NV defect electron spin, which can be efficiently polarized and read out with a long coherence time, even at room temperature.~\cite{Balasubramanian_NatMater_2009,Mizuochi_PRB_2009} \\
\indent Two nuclear spin species commonly interact with the NV defect, namely the intrinsic nitrogen atom of the defect and \carb atoms placed randomly in the diamond lattice. The hyperfine interaction between the NV defect electron spin and its nitrogen atom has been extensively characterized over the last years, both for native NV defects associated with $^{14}$N atoms and for implanted NV defects with $^{15}$N isotopes.~\cite{Felton_PRB2009} Since, the nitrogen nuclear spin occurs deterministically, it could be used as a robust memory qubit for scalable architectures of diamond-based quantum information protocols.~\cite{Fuchs_NatPhys2011} In addition, because the nitrogen nuclear spin shares its symmetry axis with the NV defect, single-shot read out measurements can be achieved at room temperature.~\cite{Neumann_Science2010} In contrast, \carb nuclear spins occur randomly in the diamond lattice and can be used as a platform for studying the coherent dynamics of multi-spin systems as well as for increasing the number of qubits in diamond-based quantum registers.\\   
\indent In this article, we report a systematic study of the hyperfine interaction between single NV defects and individual $^{13}{\rm C}$ nuclear spins. Using pulsed-ESR spectroscopy, we analyze the possible hyperfine coupling strengths induced by \carb placed at different lattice sites of the diamond matrix. Excluding the well-known $130$~MHz splitting linked to \carb placed in the nearest neighbour lattice sites of the vacancy,~\cite{Loubser,Felton_PRB2009} we focus on lattice sites corresponding to hyperfine splittings ranging from $14$~MHz to $400$~kHz. For each lattice site, properties of the \carb hyperfine tensor are inferred by using nuclear spin polarization measurements and by studying the magnetic field dependence of the hyperfine splitting. This work, which completes a recent work by Smeltzer {\it et al.},~\cite{Smeltzer_NJP_2011} is relevant for the realization of multipartite entangled state among multi-nuclear spin systems as well as for extending single shot read-out measurements recently observed on the $^{14}$N nuclear spin.~\cite{Neumann_Science2010}

\section{Single NV defects coupled with nearby nuclear spins through hyperfine interaction}

\subsection{Experimental arrangement}
\begin{figure}[h!]
\includegraphics[width = 8.7cm]{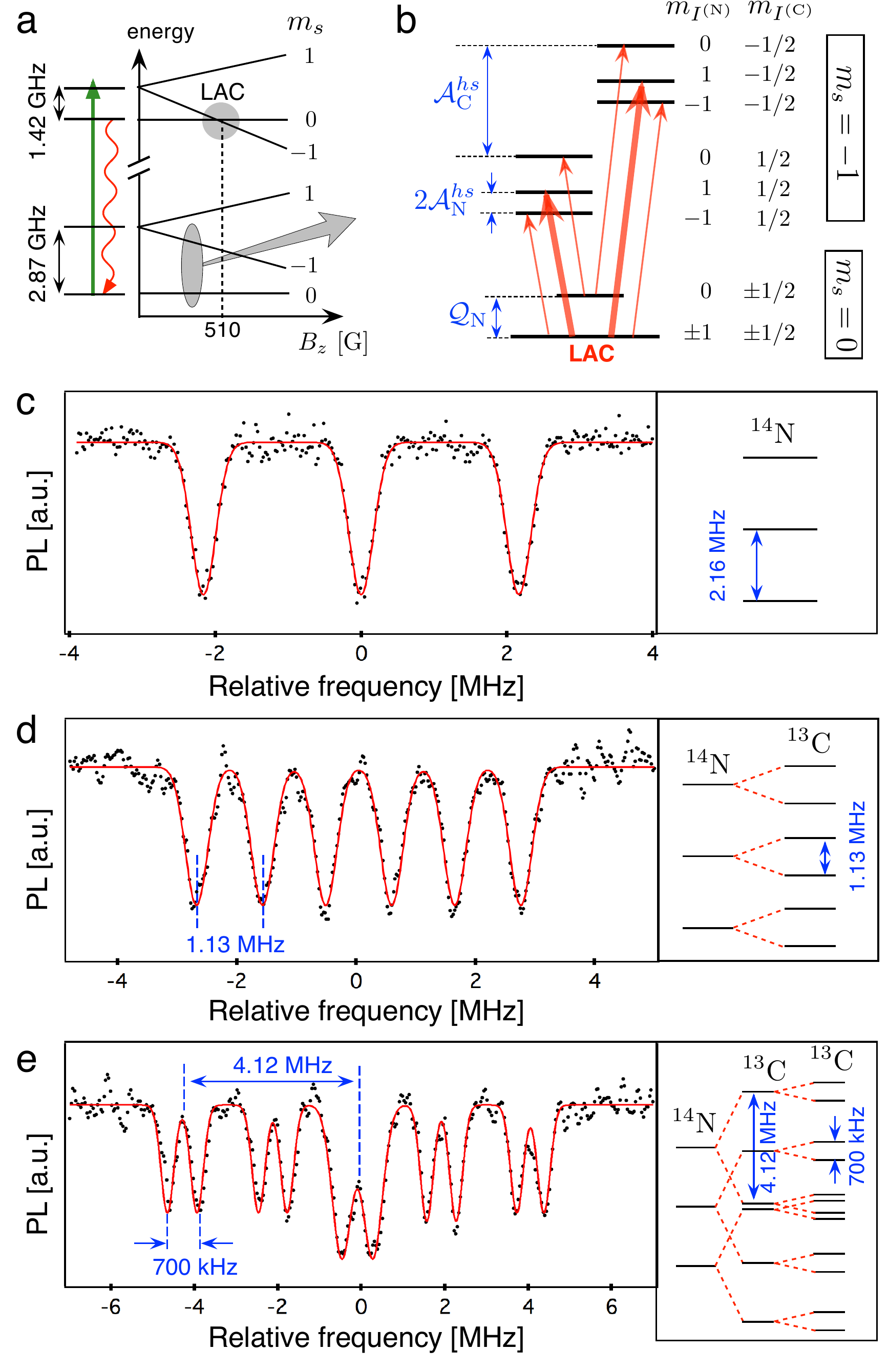}
\caption{(Color online) (a)- Energy-level diagram of the NV defect as a function of the amplitude of a static magnetic field $B$ applied along the NV defect axis. (b)-Hyperfine structure of the $m_{s}=0$ and $m_{s}=-1$ ground-state manifolds for a NV defect coupled to its $^{14}{\rm N}$ (nuclear spin projection $m_{I^{\rm (N)}}$) and a single $^{13}{\rm C}$ (nuclear spin projection $m_{I^{\rm (C)}}$). The energy level scheme is given for a positive $^{13}{\rm C}$ hyperfine splitting, {\it i.e.} $\mathcal{A}_{zz}>0$, and considering that off-diagonal components of the \carb hyperfine tensor are negligible. ESR spectra exhibit six nuclear-spin conserving transitions (red arrows). All notations are defined in the main text. (c) to (e)- ESR spectra recorded for a single NV defect coupled to its $^{14}$N and (c) zero, (d) one, and (e) two nearby $^{13}{\rm C}$. The blue values indicate the corresponding hyperfine splittings and solid lines are data fit with Gaussian functions. For this experiments, a magnetic field $B\approx 20$~G is applied along the NV defect axis in order to lift the degeneracy of the $m_{s}=\pm1$ electron spin sublevels. For a magnetic field amplitude $B\approx 510$ G, a level anti-crossing (LAC) in the excited-state (highlighted in (a)) induces nuclear-spin polarization such that only two nuclear-spin conserving transitions are observed, as indicated with bold arrows in (b) [see Fig.~\ref{Fig5}].
}
\label{Fig1}
\end{figure}

The negatively charged NV defect in diamond consists of a substitutional nitrogen atom (N) associated with a vacancy (V) in an adjacent lattice site of the diamond matrix. Its ground state is a spin triplet $S=1$ with an intrinsic spin quantization axis provided by the NV defect symmetry axis ($z$) and a zero-field splitting $D_{gs}=2.87$~GHz between 
$m_{s}=0$ and $m_{s}=\pm 1$ spin sublevels [Fig.~\ref{Fig1}(a)].~\cite{Manson_PRB2006,Maze_NJP_2011} The excited state is also a spin triplet, associated with a broadband and perfectly photostable red photoluminescence (PL), which enables optical detection of single NV defects using confocal microscopy.~\cite{Gruber_Science1997} Besides, the excited state is an orbital doublet which is averaged at room temperature,~\cite{Rogers_NJP2009,Batalov_PRL2009} leading to a zero-field splitting $D_{es}=1.42$~GHz with the same quantization axis and similar gyromagnetic ratio as in the ground state~($g_{e}\approx 2$).~\cite{Fuchs_PRL2008,Neumann_NJP2009} Radiative transition selection rules associated with the spin state quantum number provide a high degree of spin polarization in the $m_{s}=0$ sublevel through optical pumping. In addition, the PL intensity is significantly higher when the $m_{s}=0$ state is populated, allowing the detection of electron spin resonances (ESR) on a single NV defect by optical means.~\cite{Gruber_Science1997} \\
\indent We investigate native NV defects in an ultra-pure synthetic type IIa diamond crystal grown using a microwave-assisted chemical vapor deposition (CVD) process (Element 6). In such a sample with a natural abundance of $^{13}\textrm{C}$ isotope ($1.1\%$), the typical coherence time $T_{2}^{*}$ of the NV defect electron spin is on the order of a few~$\mu$s, corresponding to an inhomogeneous dephasing rate $\gamma_{2}^{*}$ on the order of a few hundreds of~kHz.~\cite{Balasubramanian_NatMater_2009,Mizuochi_PRB_2009} \\
\indent Individual NV defects are optically addressed at room temperature using a confocal microscope combined with a photon-counting detection system. A permanent magnet is used to apply a static magnetic field along the NV defect axis while ESR transitions are driven with a microwave field applied through a copper microwire directly spanned on the diamond surface. High resolution ESR spectroscopy of single NV defects coupled by hyperfine interaction with nearby $^{13}\textrm{C}$ nuclear spins is achieved through repetitive excitation of the NV defect with a resonant microwave $\pi$-pulse followed by a $300$-ns read-out laser pulse.~\cite{Dreau_PRB2011}  ESR spectra are recorded by continuously repeating this sequence while sweeping the $\pi$-pulse frequency and recording the PL intensity. Owing to spin-dependent PL of the NV defect, ESR is evidenced as a drop of the PL signal [Fig.~\ref{Fig1}(c)]. For all the experiments, the microwave power is adjusted in order to set the $\pi$-pulse duration between $2$ and $3 \ \mu$s, as verified by recording electron spin Rabi oscillations. The $\pi$-pulse duration is thus on the order of the NV defect electron-spin coherence time. In this situation, the ESR profile is Gaussian and its linewidth is given by the inhomogeneous dephasing rate $\gamma_{2}^{*}$ of the NV defect electron spin, since power broadening is fully cancelled in the measurement.~\cite{Dreau_PRB2011}

\subsection{Spin Hamiltonian}

\indent We consider single NV defects associated with native $^{14}$N isotopes ($99.6\%$ abundance), corresponding to a nuclear spin $I^{\rm (N)}=1$. For a magnetic field $B$ applied along the NV defect axis, the ground-state spin Hamiltonian in frequency unit reads as
\begin{equation}
\mathcal{H}_{0}=D_{gs}\hat{S}_{z}^{2}+\gamma_{e}B\hat{S}_{z}+\gamma_{n}^{\rm (N)}B\hat{I}_{z}^{\rm (N)}+\mathcal{Q}_{\rm N}(\hat{I}_{z}^{\rm (N)})^{2}+\mathbf{\hat{S}}\cdot{\mathcal A}_{\rm N}\cdot\mathbf{\hat{I}^{(N)}}
\label{eq:h}
\end{equation}
where $\mathcal{Q}_{\rm N}=-5.01$~MHz is the $^{14}$N quadrupole splitting,~\cite{Felton_PRB2009} $\mathcal{A}_{\rm N}$ its hyperfine tensor and $\gamma_{e}$ (resp. $\gamma_{n}^{\rm (N)}$) is the electron spin (resp.~$^{14}$N nuclear spin) gyromagnetic ratio. The $^{14}$N hyperfine tensor has been extensively characterized over the last years and leads to a splitting of $\mathcal{A}_{\rm N}^{\rm hs}=-2.16$ MHz between ESR frequencies associated with different $^{14}$N nuclear spin projections~\cite{Steiner_PRB2010}. For single NV defects without any strongly coupled $^{13}$C nuclear spins, the ESR spectrum thus always exhibit three hyperfine lines [Fig.~\ref{Fig1}(b) and (c)].\\
\indent When a neighbouring lattice site of the NV defect is occupied by a $^{13}$C isotope, corresponding to a nuclear spin $I^{\rm (C)}=1/2$, the spin Hamiltonian becomes
\begin{equation}
\mathcal{H}=\mathcal{H}_{0}+\gamma_{n}B\hat{I}_{z}^{\rm (C)}+\mathbf{\hat{S}}\cdot{\mathcal A}_{C}\cdot\mathbf{\hat{I}^{\rm (C)}}
\label{eq:h}
\end{equation}
where $\gamma_{n}$ is the gyromagnetic ratio of the \carb nuclear spin and $\mathcal{A}_{C}$ its hyperfine tensor defined by 
\begin{equation}
\mathcal{A}_{C}=\left(\begin{array}{ccc}\mathcal{A}_{xx} & \mathcal{A}_{xy} & \mathcal{A}_{xz} \\\mathcal{A}_{yx} & \mathcal{A}_{yy} & \mathcal{A}_{yz} \\
\mathcal{A}_{zx} & \mathcal{A}_{zy} & \mathcal{A}_{zz}\end{array}\right) \\ .
\end{equation}
In the secular approximation, {\it i.e.} neglecting $\hat{S}_{x}$ and $\hat{S}_{y}$ terms, this Hamiltonian simplifies as :
\begin{equation}
\mathcal{H}=\mathcal{H}_{0}+\gamma_{n}B\hat{I}_{z}^{\rm (C)}+\hat{S}_{z}\sum_{i}{\mathcal{A}}_{zi}\hat{I}_{i}^{\rm (C)} \ .
\label{Hamilto1}
\end{equation}
\indent For clarity purpose, we restrict the study to the $m_{s}=0$ and $m_{s}=-1$ electronic spin manifolds with a fixed $^{14}$N nuclear spin projection, {\it e.g.} $m_{I^{\rm (N)}}=+1$. All the results are identical while considering the two other $^{14}$N nuclear spin projections. In the basis $\left\{\left|m_{s},m_{I^{\rm (C)}} \right.\rangle\right\}=\left\{\left|0,\frac{1}{2} \right.\rangle ; \left|0,-\frac{1}{2} \right.\rangle ; \left|-1,\frac{1}{2} \right.\rangle ; \left|-1,-\frac{1}{2} \right.\rangle \right\}$ and considering $\mathcal{A}_{zz}>0$, the Hamiltonian described by equation~(\ref{Hamilto1}) can be written as 
\begin{equation}
\mathcal{H}=\left[\begin{array}{cccc}\vspace{0.2cm}
\frac{\gamma_{n}B}{2} & 0 & 0 & 0\\ \vspace{0.2cm}
0 & -\frac{\gamma_{n}B}{2}  & 0 & 0\\ \vspace{0.2cm}

0 & 0 & \Sigma-\frac{\mathcal{A}_{zz}-\gamma_{n}B}{2} & -\frac{\mathcal{A}_{\rm nd}}{2}e^{-i\phi} \\ \vspace{0.2cm}

0 & 0 & -\frac{\mathcal{A}_{\rm nd}}{2}e^{i\phi} & \Sigma+\frac{\mathcal{A}_{zz}-\gamma_{n}B}{2}
\end{array}\right] \ ,
\label{Hamilto}
\end{equation}
where 
\begin{equation}
\left\{\begin{array}{lll} \vspace{0.2cm}
\mathcal{A}_{\rm nd}&=&\sqrt{\mathcal{A}_{xz}^{2}+\mathcal{A}_{yz}^{2}}\\ \vspace{0.2cm}
\Sigma&=&D_{gs}-\gamma_{e}B-\mathcal{A}_{N}^{hs}\\
\tan \phi&=&\displaystyle\frac{\mathcal{A}_{yz}}{\mathcal{A}_{xz}} \ .
\end{array}
\right.
\end{equation}
We note that the $^{14}$N nuclear Zeeman splitting and the $^{14}$N quadrupole splitting are ignored in Eq.(\ref{Hamilto}) since they only introduce a global shift of the energy levels. In this framework, the eigenenergies of the NV defect electron spin coupled by hyperfine interaction with a single $^{13}$C nuclear spin are given by
\begin{eqnarray}
\nu_{1}&=&+\frac{\gamma_{n}B}{2}\\
\nu_{2}&=&-\frac{\gamma_{n}B}{2}\\
\nu_{3}&=&\Sigma-\frac{1}{2}\sqrt{\mathcal{A}_{\rm nd}^{2}+({\mathcal{A}}_{zz}-\gamma_{n}B)^{2}}\\
\nu_{4}&=&\Sigma+\frac{1}{2}\sqrt{\mathcal{A}_{\rm nd}^{2}+({\mathcal{A}}_{zz}-\gamma_{n}B)^{2}} 
\end{eqnarray}
and the associated eigenstates can be written as 
\begin{eqnarray*}
\left|\psi_{1}\right.\rangle&=&\left|0,\frac{1}{2} \right.\rangle\\
\left|\psi_{2}\right.\rangle&=&\left|0,-\frac{1}{2} \right.\rangle\\
\left|\psi_{3}\right.\rangle&=&\left|+\right.\rangle\\
&=&\cos\left(\frac{\theta}{2}\right)\left|-1,\frac{1}{2} \right.\rangle+\sin\left(\frac{\theta}{2}\right)e^{i\phi}\left|-1,-\frac{1}{2} \right.\rangle\\
\left|\psi_{4}\right.\rangle&=&\left|- \right.\rangle\\
&=&-\sin\left(\frac{\theta}{2}\right)e^{-i\phi}\left|-1,\frac{1}{2} \right.\rangle+\cos\left(\frac{\theta}{2}\right)\left|-1,-\frac{1}{2} \right.\rangle
\end{eqnarray*}
where
\begin{equation}
\tan\theta=\frac{\mathcal{A}_{\rm nd}}{\mathcal{A}_{zz}-\gamma_{n}B} \ .
\label{teta}
\end{equation}
\indent Allowed ($T_{13},T_{24}$) and forbidden ($T_{14},T_{23}$) electron spin transition frequencies are therefore given by 
\begin{eqnarray}
T_{13}&=&\Sigma-\frac{1}{2}\sqrt{\mathcal{A}_{\rm nd}^{2}+({\mathcal{A}}_{zz}-\gamma_{n}B)^{2}}-\frac{\gamma_{n}B}{2}\\
T_{24}&=&\Sigma+\frac{1}{2}\sqrt{\mathcal{A}_{\rm nd}^{2}+({\mathcal{A}}_{zz}-\gamma_{n}B)^{2}}+\frac{\gamma_{n}B}{2}\\
T_{14}&=&\Sigma+\frac{1}{2}\sqrt{\mathcal{A}_{\rm nd}^{2}+({\mathcal{A}}_{zz}-\gamma_{n}B)^{2}}-\frac{\gamma_{n}B}{2}\\
T_{23}&=&\Sigma-\frac{1}{2}\sqrt{\mathcal{A}_{\rm nd}^{2}+({\mathcal{A}}_{zz}-\gamma_{n}B)^{2}}+\frac{\gamma_{n}B}{2}\ ,
\label{Transit}
\end{eqnarray}
where the numbering is determined by the corresponding energy levels, $T_{ik}=\nu_{k}-\nu_{i}$.\\
\indent If $|\mathcal{A}_{zz}-\gamma_{n}B|$ and $\mathcal{A}_{\rm nd}$ have the same order of magnitude, four electron spin transitions can thus be observed for each $^{14}$N nuclear spin projection, with a relative intensity between forbidden and allowed transitions given by $\tan^{2}(\frac{\theta}{2})$. When $\mathcal{A}_{zz}=\gamma_{n}B$, {\it i.e.} when $\theta=\pi /2$, the amplitudes of allowed and forbidden electron spin transitions are identical. This particular case is experimentally observed and analysed in Section~\ref{OffDiag} [Fig.~\ref{Fig3} and Fig.~\ref{Fig4}].\\
\indent In the limit where $|\mathcal{A}_{zz}-\gamma_{n}B|\gg \mathcal{A}_{\rm nd}$, {\it i.e.} when the angle $\theta$ approaches zero, then 
\begin{eqnarray*}
\left|\psi_{1}\right.\rangle&=&\left|0,\frac{1}{2} \right.\rangle\\
\left|\psi_{2}\right.\rangle&=&\left|0,-\frac{1}{2} \right.\rangle\\
\left|\psi_{3}\right.\rangle&=&\left|-1,\frac{1}{2} \right.\rangle\\
\left|\psi_{4}\right.\rangle&=&\left|-1,-\frac{1}{2} \right.\rangle \ .
\end{eqnarray*}
In this situation, the $^{13}$C nuclear spin projection along the NV defect axis is identical in the $m_{s}=0$~and $m_{s}=-1$ electronic spin manifolds and only two nuclear-spin conserving transitions can be observed ($T_{13}$ and $T_{24}$). By considering all the $^{14}$N nuclear spin projections, the ESR spectrum thus exhibits six nuclear-spin conserving transitions as schematically depicted in Figure~\ref{Fig1}(b).

\section{Experimental observations}
\subsection{Low magnetic field hyperfine splittings}
\begin{figure}[t]
\includegraphics[width=8.9cm]{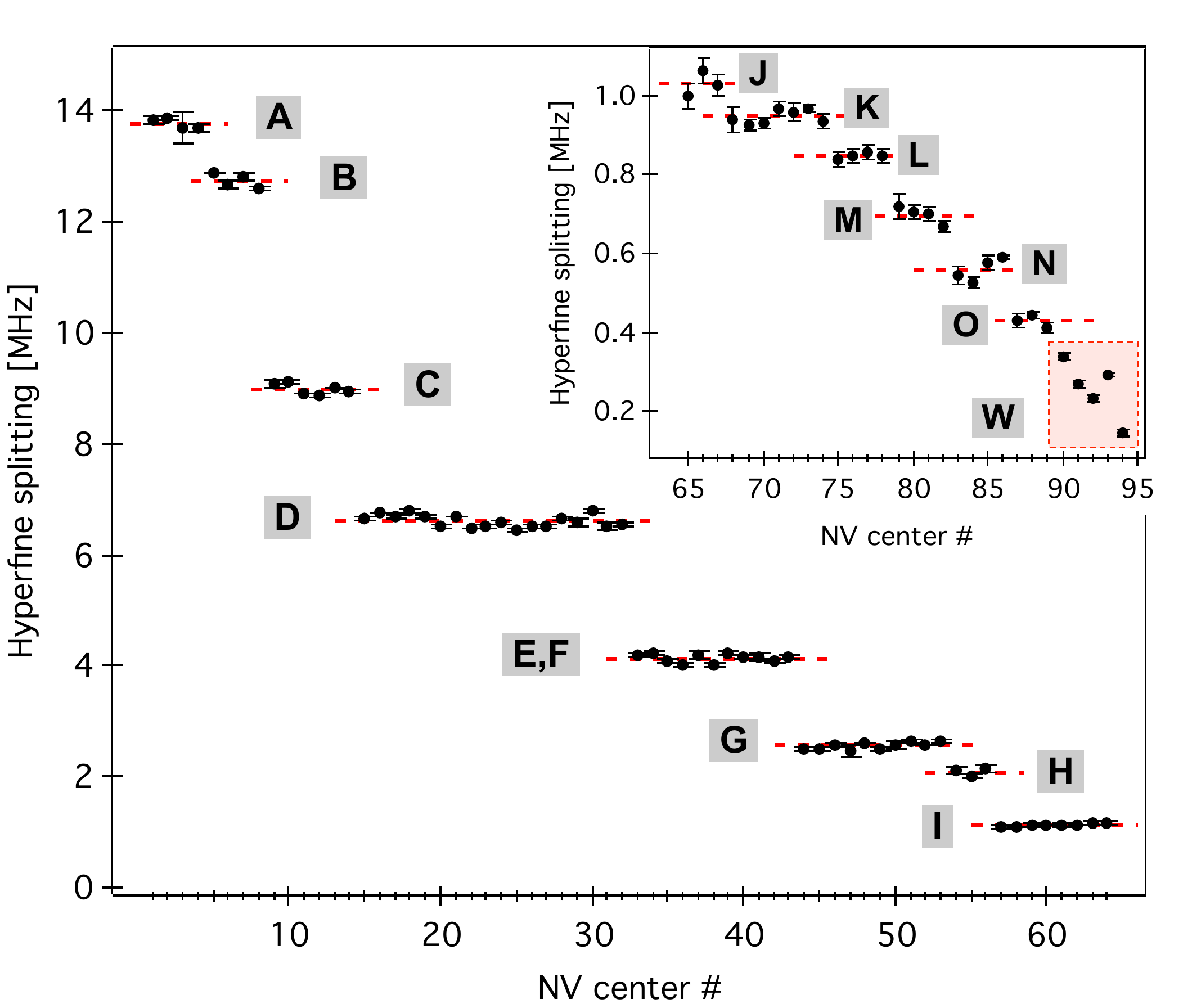}
\caption{Distribution of \carb hyperfine splittings observed for a set of roughly 400 single NV defects at low magnetic field ($B\approx 20$~G). The well-documented~\cite{Loubser,Felton_PRB2009} $130$ MHz splitting linked to a \carb placed in the nearest neighbour lattice sites of the vacancy is not represented. The capital letters denote the lattice sites corresponding to a given hyperfine splitting. The dash red lines represent the mean value of each \carb family (A to O) and error bars represent estimated errors in ESR spectra fit with a 95 $\%$ confidence interval. For a given \carb family, the small dispersion of the hyperfine splittings is attributed to local paramagnetic defects, which could shift the nuclear spin levels.The mean of the hyperfine splitting associated with each \carb family are summarized in table \ref{table_splittings}.
}
\label{Fig2}
\end{figure}
\indent We first study the possible values of hyperfine splittings induced by nearby $^{13}$C nuclear spins for a low magnetic field magnitude ($B\approx 20$~G) applied along the NV defect axis. In this condition, the nuclear Zeeman term can be neglected ($T_{13}=T_{23}$, $T_{24}=T_{14}$) and the $^{13}$C hyperfine interaction leads to a splitting 
$$\mathcal{A}_{\rm C}^{hs}(@ 20 \ {\rm G}) =T_{24}-T_{13}=\pm\sqrt{\mathcal{A}_{\rm nd}^{2}+{\mathcal{A}}_{zz}^{2}}$$ between ESR frequencies associated with different $^{13}$C nuclear spin projections, with a sign given by the one of the $\mathcal{A}_{zz}$ component [Fig.~\ref{Fig1}(b)].   \\
\indent A systematic study of the $^{13}$C hyperfine coupling strength was realized by performing pulsed-ESR spectroscopy on a set of roughly $400$ single NV defects. Even though most of ESR spectra only show the three hyperfine lines linked to the $^{14}$N nuclear spin~[Fig.~\ref{Fig1}(c)], the subset ($\approx 25\%$) of NV defects strongly coupled to one (resp. two) \carb nuclear spin exhibit six (resp. twelve) resonance lines~[Fig.~\ref{Fig1}(d)-(e)]. From each ESR spectrum, the hyperfine splitting between the \carb manifolds was extracted from a fit with Gaussian functions, constrained to have the same linewidth and amplitude, and with a fixed splitting between $^{14}$N hyperfine lines ($\mathcal{A}_{\rm N}^{\rm hs}=-2.16$ MHz).
\begin{table}[b]
\caption{\label{table_splittings}Summary of hyperfine splittings $\mathcal{A}_{\rm C}^{hs}$ measured at $B=20$~G and at $B=510$~G. The last column indicates for each \carb family the polarization efficiency $\mathcal{P}$ measured at the excited-state LAC using pulsed-ESR spectrocopy, {\it i.e.} with an optical pumping duration of $300$~ns. Numbers between parenthesis indicate the standard deviation on the last digit for a number of NV centers that is given between square brackets.}
\begin{ruledtabular}
\begin{tabular}{cccc}
 \carb family &$\mathcal{A}_{\rm C}^{hs}(@ \ 20 \ {\rm G})$&$\mathcal{A}_{\rm C}^{hs}(@ \ 510 \ {\rm G})$& $\mathcal{P}$ [$\%$]\\
  &[MHz]&[MHz]&\\
\hline
	 A & 13.78 (9) [4] & 13.69 (5) [3] &  52 (20) [3] \\ 
    		B & 12.8 (1) [4]  &  12.73 (1) [3]  &  53 (13) [3]\\ 
    		C & -9.0 (1) [6] &  -8.9 (2) [4]  & -56 (8) [4]  \\ 
   		D & -6.6 (1) [18]  & -6.55 (2) [6] & -65 (10) [6] \\
    		E, F & 4.12 (8) [11]  & 4.21 (8) [3] &  43 (6) [3]\\ 
    		G & 2.55 (6) [10]  & 2.54 (3) [4]  & 34 (6) [4]\\ 
    		H & 2.09 (8) [3] & 2.15 (4) [3]  & 54 (4) [3]\\ 
    		I & 1.13 (2)  [8]  & 1.20 (6) [2]  &  0 (1) [2] \\ 
    		J & -1.03 (3) [3]  & -0.99 (1) [2]  & -3 (1) [2] \\ 
    		K & 0.95 (2) [7] & 0.92 (7) [3]  & 60 (8) [3]\\ 
    		L & 0.85 (1) [4]  & 0.86 (3) [3] & 1 (1) [3]\\ 
    		M & -0.70 (3) [4] & -0.69 (2) [3]  & -4 (4) [3]\\ 
		N & 0.56 (3) [4] & 0.52 (2)  [2] &  2 (4) [2] \\ 
    		O & 0.43 (2) [3] & 0.40 (2) [3] &  13 (7) [3]\\ 
\end{tabular}
\end{ruledtabular}
\end{table}

\indent As shown in Figure~\ref{Fig2}, the \carb hyperfine splittings are distributed over a set of discrete values corresponding to different lattice sites of the diamond matrix. Following the notations introduced in Ref. \cite{Smeltzer_NJP_2011}, hyperfine splittings such that $\mathcal{A}_{\rm C}^{hs}>2$~MHz are labelled from A to H, corresponding to possible lattice sites inferred from {\it ab initio} calculations~\cite{Gali_PRB_2008,Gali_PRB_2009}. Smaller hyperfine splitting are arbitrarily denoted from I to O, but no correspondence with a specific lattice site of the diamond matrix is currently available, since the accuracy of {\it ab initio} calculation is hardly better than a few hundred of kHz. We note that \carb families are sorted according to the strength of the hyperfine coupling and by their polarization response at the excited-state level anti-crossing (LAC), as analysed in details in Section~\ref{Polar}.\\
\indent For a few single NV defects, the inhomogeneous dephasing rate $\gamma_{2}^{*}$ was small enough to detect hyperfine splittings below $300$~kHz, labelled W in Figure~\ref{Fig2}. Within our statistic, it is not possible to clearly identify common lattice sites for such weak couplings. However, the possible discrete values of the hyperfine splittings should reach a quasi-continuum when the \carb is placed far away from the NV defect because many different lattice sites might exhibit similar hyperfine coupling strengths. A more systematic study of weaker \carb hyperfine splittings could be pursued by using a CVD-grown diamond isotopically enriched with $^{12}\textrm{C}$ atoms, in which the inhomogeneous dephasing rate of the NV defect electron spin can reach a few tens of kHz.~\cite{Balasubramanian_NatMater_2009,Mizuochi_PRB_2009} The drawback would obviously be a lower rate of detection of \carb nuclear spin per single NV defect.\\
\indent The mean of the hyperfine splitting associated with each \carb family is summed up in table \ref{table_splittings}. We note that for families A to G, the results are in good agreement with recent single site measurements.\cite{Smeltzer_NJP_2011} 

\subsection{Effect of off-diagonal components of the hyperfine tensor}
\label{OffDiag}
\begin{figure}[t]
\includegraphics[width=8.5cm]{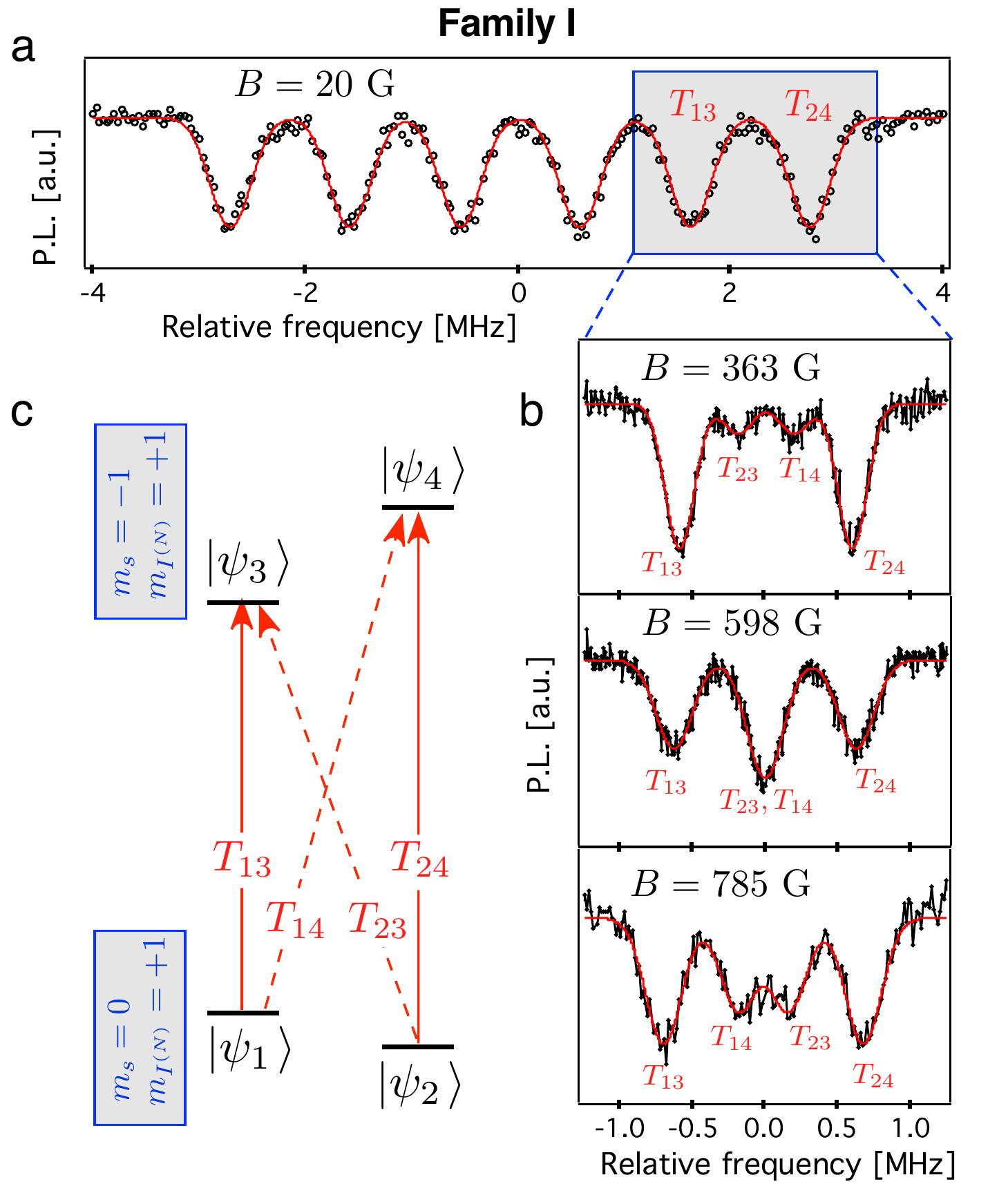}
\caption{(a)-Typical ESR spectrum recorded at low magnetic field ($B=20$~G) for a single NV defect coupled with a single \carb nuclear spin belonging to family I ($\mathcal{A}_{\rm C}^{hs}=1.12$~MHz). (b)-Zoom on the hyperfine transitions associated with a $^{14}$N nuclear spin projection $m_{I^{\rm (N)}}=+1$. Forbidden electron spin transitions ($T_{23}$,$T_{14}$) appear in the ESR spectrum when the magnetic field magnitude is increased. (c)-Energy level structure of the $m_{s}=0$ and $m_{s}=-1$ ground-state manifolds for a $^{14}$N nuclear spin projection $m_{I^{\rm (N)}}=1$. Solid (resp. dashed) arrows indicate allowed (resp. forbidden) electron spin transitions. All notations are defined in the main text.}
\label{Fig3}
\end{figure}
With the aim of observing the effects of anisotropic components of the hyperfine tensor, the \carb coupling strengths were first measured while applying a magnetic field near $B=510$~G along the NV defect axis. In this situation, the hyperfine splitting between allowed electron spin transitions ($T_{13}$ and $T_{24}$) is given by 
$$\mathcal{A}_{\rm C}^{hs}(@ 510 \ {\rm G})=\sqrt{\mathcal{A}_{\rm nd}^{2}+({\mathcal{A}}_{zz}-\gamma_{n}B)^{2}}+\gamma_{n}B \ .$$
\indent As summarized in table \ref{table_splittings}, most \carb hyperfine splittings are identical within the error bars at low and high magnetic fields. However, for some \carb lattice sites, {\it e.g.} for families E and I, their relative differences becomes significant, which is an indication that anisotropic components of the hyperfine tensor lead to a different nuclear spin projection along the NV defect axis in the $m_s~=~0$~and $m_{s}=-1$ electronic spin manifolds.\\
\begin{figure}[t]
\includegraphics[width=9cm]{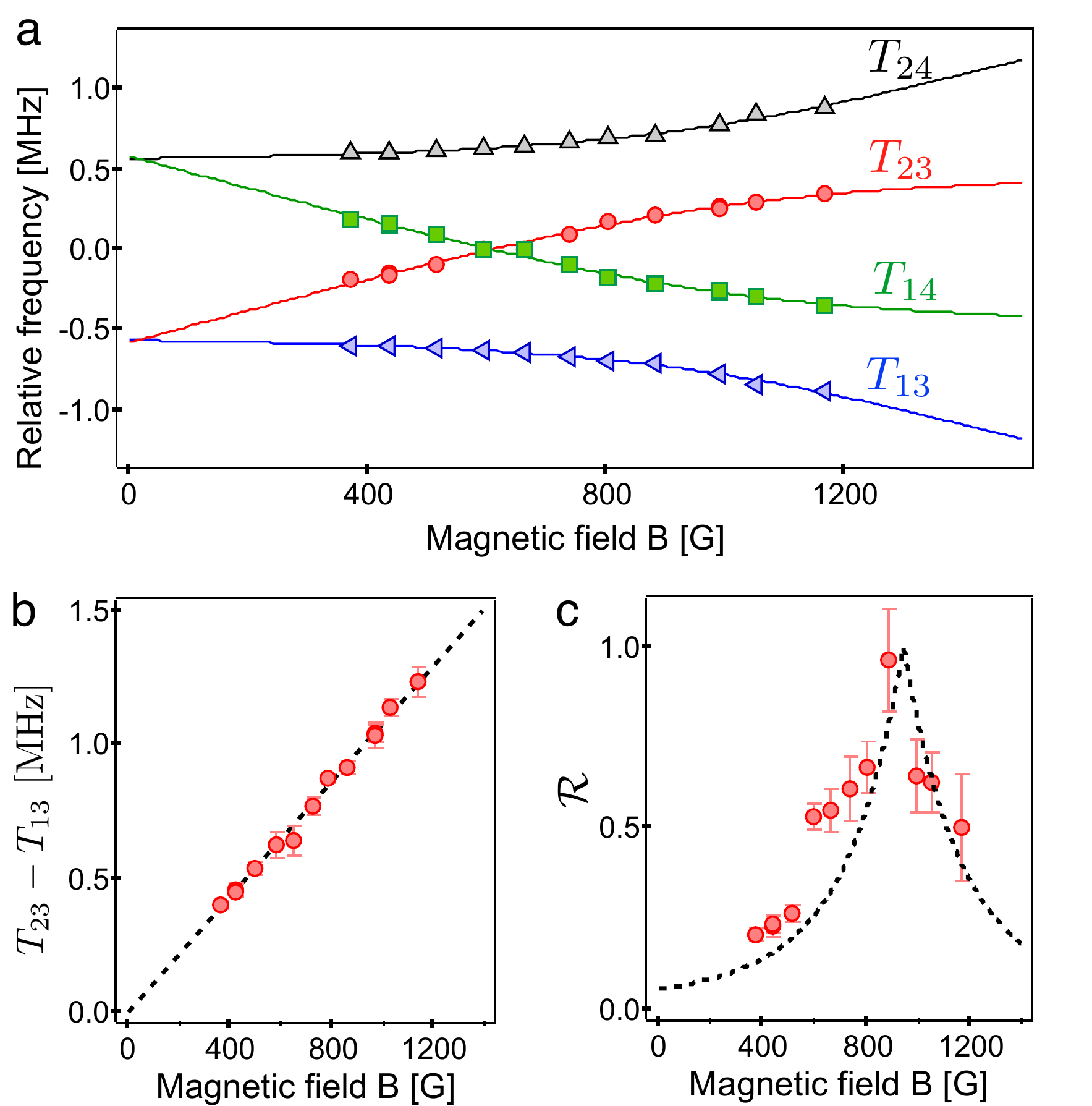}
\caption{(a)-Frequencies of forbidden ($T_{23}$,$T_{14}$) and allowed ($T_{13}$,$T_{24}$) electron spin transitions as a function of the magnetic field magnitude. Solid lines are data fits using Eqs.~(12) to (15) with $\mathcal{A}_{zz}$ and $\mathcal{A}_{\rm nd}$ as fitting parameters.(b)-$T_{23}-T_{13}$ as a function of the magnetic field magnitude. The black dashed line is data fit with a linear function, from which the \carb nuclear spin gyromagnetic ratio $\gamma_{n}=1.07\ (1)$~kHz.G$^{-1}$ is extracted. (c)-Relative intensity $\mathcal{R}$ between forbidden and allowed electron spin transitions as a function of the magnetic field. The dashed line is the theoretical expectation $\mathcal{R}=\tan^{2}(\frac{\theta}{2})$, where $\theta$ is defined by Eq.~(\ref{teta}) with $\mathcal{A}_{zz}=1.02$~MHz and $\mathcal{A}_{\rm nd}=0.51$~MHz. The slight deviation of experimental data points is attributed to an imperfect microwave $\pi$-pulse in the pulsed-ESR spectroscopy scheme.}
\label{Fig4}
\end{figure}
\indent In the limit where the components $|\mathcal{A}_{zz}-\gamma_{n}B|$ and $\mathcal{A}_{\rm nd}$ reach the same order of magnitude, forbidden electron spin transitions ($T_{14}$ and $T_{23}$) can even be identified in the ESR spectrum. This situation is observed for single NV defects coupled with \carb belonging to family I, as depicted in Figure~\ref{Fig3}. At low magnetic field, since $T_{13}=T_{23}$ and $T_{24}=T_{14}$, only two \carb hyperfine transitions can be observed for each $^{14}$N nuclear spin projection [Fig.~\ref{Fig3}(a)]. When the magnetic field magnitude increases, forbidden transitions appears in the ESR spectrum [Fig.~\ref{Fig3}(b)-(c)]. The frequencies of forbidden and allowed electron spin transitions were measured as a function of the magnetic field magnitude [Fig.~\ref{Fig4}(a)]. Data fitting with Eqs.~(12) to (15) allows to extract $\mathcal{A}_{zz}=1.02 \ (2)$~MHz and $\mathcal{A}_{\rm nd}=0.51 \ (2)$~MHz. In addition, plotting $T_{23}-T_{13}$ as a function of the magnetic field allows measurement of the \carb nuclear spin gyromagnetic ratio $\gamma_{n}=1.07\ (1)$~kHz.G$^{-1}$ [Fig.~\ref{Fig4}(b)], which is in perfect agreement with the value inferred from nuclear magnetic resonance measurements.~\cite{Handbook} \\
\indent The relative intensity $\mathcal{R}$ between forbidden and allowed electron spin transitions was estimated by measuring the ratio between the integral of the ESR line at frequency $T_{23}$ and  $T_{13}$ [Fig.~\ref{Fig4}(c)]. As expected $\mathcal{R}\approx 1$ when $\mathcal{A}_{zz}=\gamma_{n}B$, {\it i.e.} when $B\approx 950$~G. In this situation, $\theta=\pi/2$ and the \carb nuclear quantization axis is perpendicular to the NV defect axis in the $m_{s}=-1$ electron spin manifold.\\
\indent We note that this system provides an efficient $\Lambda$-scheme that could be used for coherent population trapping of a single nuclear spin and electromagnetically induced transparency experiments in the microwave domain.~\cite{Togan_Nature2011}

\subsection{Polarization efficiency through optical pumping at the excited-state LAC}
\label{Polar}
\begin{figure}[h!]
\includegraphics[width=9cm]{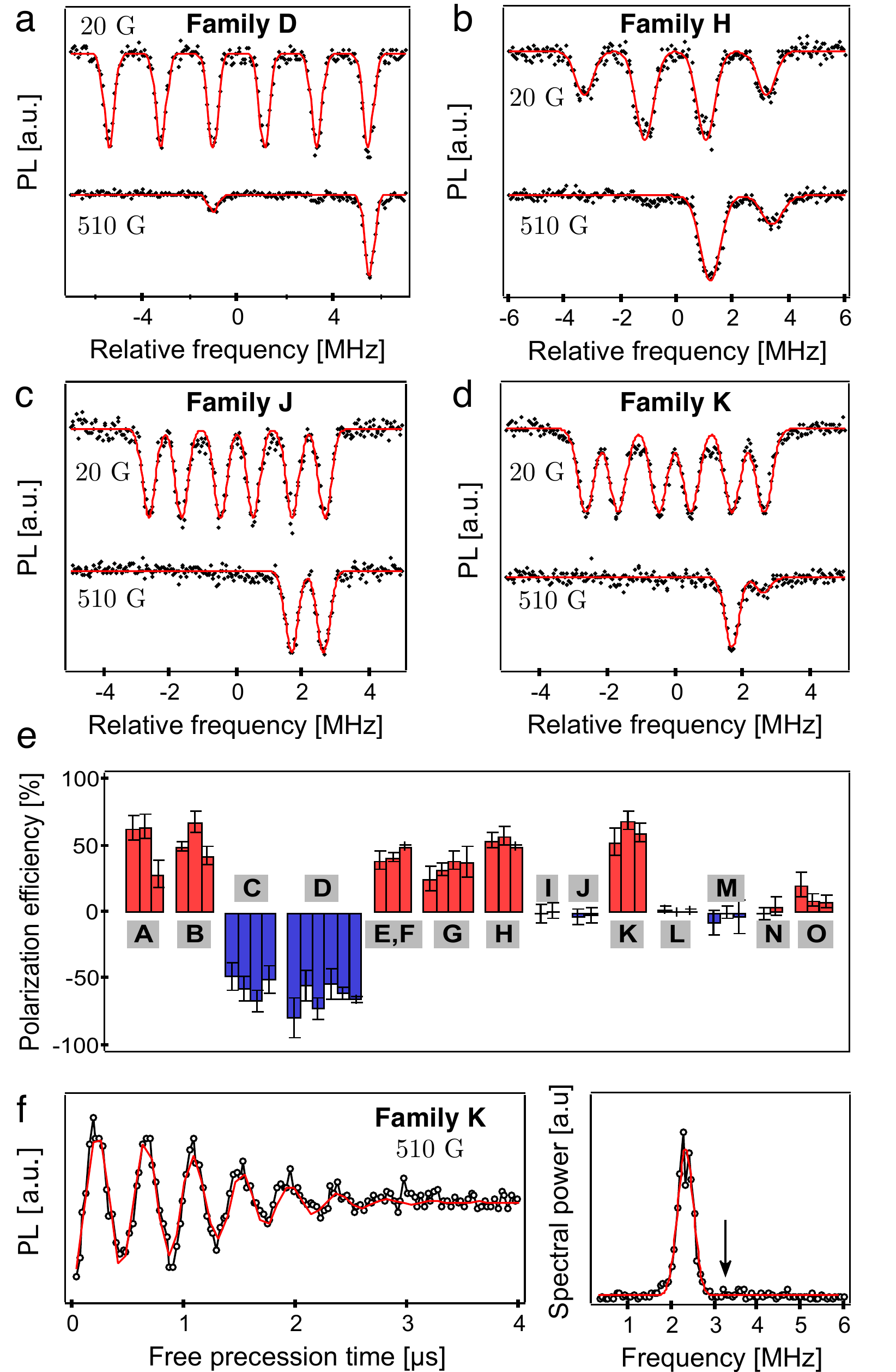}
\caption{(a) to (d)-Typical pulsed-ESR spectra recorded at low magnetic field ($B\approx 20$~G) and at the excited-state LAC ($B\approx 510$~G) for single NV defects coupled by hyperfine interaction with a single \carb belonging to family (a) D, (b) H, (c) J, and (d) K. The $^{14}$N nuclear spin is perfectly polarized while the polarization efficiency of nearby \carb strongly depends on the lattice site. (e)-Polarization efficiency of a set of single NV defect coupled with different families of $^{13}$C. We note that this results are obtained by using pulsed-ESR spectroscopy, {\it i.e.} with an optical pumping duration of $300$~ns. Positive polarization (resp. negative) indicates a positive (resp. negative) hyperfine splitting. Within each \carb family, the variation of polarization efficiency is attributed to misalignment of the magnetic field along the NV defect axis. The mean polarization efficiency for each \carb family in given in Table~\ref{table_splittings}. (f)-Free induction decay recorded for the same NV defect as in (d) at the excited-state LAC and for an optical pumping duration of $3$~$\mu$s. A single line is observed in the Fourier transform, which is the signature of nearly perfect polarization of the \carb nuclear spin. }
\label{Fig5}
\end{figure}
\indent Hyperfine interactions with nearby \carb were further investigated by measuring their polarization efficiency at the excited-state LAC, while applying a static magnetic field near $510$ G along the NV axis~[Fig.~\ref{Fig1}(a)]. In this configuration, electron-nuclear-spin flip-flops mediated by hyperfine interaction in the excited-state can lead to an efficient polarization of nearby nuclear spins in their highest spin projection along the excited-state electron spin quantization axis ($z$).~\cite{Jacques_PRL2009,Childress_PRA2009,Steiner_PRB2010} Provided that the nuclear spin projection along the $z$ axis is identical in the ground state and in the excited state, nuclear-spin polarization is transferred to the ground state by non-radiative inter-system crossing through metastable singlet states responsible for electron spin polarization.~\cite{Gali_PRB_2009} The efficiency of such a process is linked to several parameters. First, the polarization $\mathcal{P}$ is reduced if the nuclear spin quantization axis differs in the ground and in the excited state, as expected for a hyperfine tensor with different principal axis in the ground and in the excited states.~\cite{Gali_PRB_2009} For a nuclear spin that possesses an identical quantization axis, the polarization efficiency is also decreased if this axis is not the same as the electron spin quantization axis, {\it i.e.} if anisotropic components of the hyperfine tensor play a significant role in the nuclear-spin dynamics, as observed in the previous section for \carb of family~I.\\
\indent Since the $^{14}$N nuclear spin quantization axis is parallel to the NV defect axis, perfect polarization is achieved in the $m_{I^{\rm (N)}}=1$ sublevel through optical pumping at the excited-state LAC.~\cite{Jacques_PRL2009,Childress_PRA2009,Steiner_PRB2010} The ESR spectrum of a single NV defect coupled with a single \carb nuclear spin thus only exhibits two hyperfine lines at the excited-state LAC, associated with the two different $^{13}$C nuclear spin projections [Fig.~\ref{Fig1}(b) and Fig.~\ref{Fig5}(a) to (d)]. For each family of \carb, the polarization efficiency $\mathcal{P}$ was measured as the difference between the integral of each \carb hyperfine lines divided by their sum. We note that the amplitude of ESR dips is also affected by the difference in electron spin readout contrast for the two \carb nuclear spin orientations when spin flip-flop processes occur in the excited-state.~\cite{Steiner_PRB2010}\\
\indent Since all possible \carb lattice sites have different hyperfine interactions with the NV electron spin, each \carb family exhibits different polarization behavior [Fig.~\ref{Fig5}(a) to (e)]. For hyperfine splittings larger than $2$~MHz (families A to H), only partial nuclear spin polarization is achieved, which qualitatively demonstrates that the nuclear spin quantization axis might differ in the ground and in the excited state, as predicted by {\it ab initio} calculations.~\cite{Gali_PRB_2008,Gali_PRB_2009} In addition, the nuclear spin quantization axis is not the same as the electron spin axis because none of these \carb families correspond to a lattice site along the NV defect axis.~\cite{Smeltzer_NJP_2011} Beyond documenting qualitatively the relative orientation of the \carb, nuclear-spin polarization measurements also indicate the sign of the hyperfine coupling. Indeed, nuclear-spin polarization into the lowest (resp. highest) \carb hyperfine frequency indicates that the hyperfine splitting is positive (resp. negative), which corresponds to a lattice site where the NV defect electron spin density is positive (resp. negative) [Fig.~\ref{Fig5}(a) and (b)]. \\
\indent For \carb with a hyperfine splitting smaller than $2$ MHz~[Fig.~\ref{Fig5}(c) and (e)], almost no polarization is observed, except for family K for which a high level of polarization ($\approx 60\%$) is evidenced in pulsed-ESR spectra [Fig.~\ref{Fig5}(d)]. However, for such a weak hyperfine interaction ($0.95$~MHz), the polarization efficiency can be limited by the optical pumping duration, since the probability of electron-nuclear-spin flip-flop per optical cycle is given by the strength of the excited-state hyperfine interaction. In the pulsed-ESR spectroscopy scheme, the optical pumping duration is set at $300$~ns in order to optimize the ESR contrast,~\cite{Dreau_PRB2011} which could be too short to reach the steady state of nuclear spin populations. In order to check if nuclear-spin polarization can be improved while increasing the optical pumping duration, the polarization efficiency was estimated by using the Fourier transform of free-induction decay (FID) measurements. For that purpose, Ramsey fringes were recorded by using the standard sequence consisting of a $3$-$\mu$s laser pulse, used both for optical pumping and spin-state readout, followed by two microwave $\pi/2$-pulses separated by a variable free evolution duration $\tau$. A typical FID signal recorded at the excited-state LAC for a single NV defect coupled with a \carb belonging to family K is shown in Figure~\ref{Fig5}(f). The Fourier transform of the FID shows a single line, which is the signature that the \carb nuclear spin is fully polarized. This result might indicate that (i) the hyperfine tensor has an identical principal axis in the ground state and in the excited state, and (ii) that the nuclear-spin quantization axis is parallel to the electron spin. Family K could thus correspond to a \carb placed on a lattice site along the NV defect axis, which shares the symmetry of the NV defect. Such \carb could be an interesting candidate for extending single shot read-out measurements recently observed on the $^{14}$N nuclear spin at room temperature.~\cite{Neumann_Science2010}\\

\section{Conclusion}
Using pulsed-ESR spectroscopy, we have reported a systematic study of the hyperfine interaction between the electron spin of a single NV defect in diamond and nearby \carb nuclear spins. We have isolated a set of discrete values of the hyperfine coupling strength ranging from 14 MHz to 400 kHz and corresponding to \carb nuclear spins placed at different lattice sites of the diamond matrix. Nuclear-spin polarization measurements suggest that one of the reported hyperfine splittings corresponds to a \carb placed on a lattice site along the NV defect axis. This work provides important information for the development of nuclear-spin based quantum registers in diamond. In addition, the set of \carb hyperfine splittings combined with their polarization behaviors, can be used for checking {\it ab initio} calculations of the NV defect spin density.~\cite{Smeltzer_NJP_2011,Gali_PRB_2008} 

\begin{acknowledgments}
The authors acknowledge L.~Childress, P.~Bertet, R.~Hanson, and F.~Jelezko for fruitful discussions. This work was supported by C'Nano \^Ile-de-France and the Agence Nationale de la Recherche (ANR) through the projects D{\sc iamag}, A{\sc dvice} and Q{\sc invc}. J.R.M. acknowledges support from Conicyt Fondecyt, Grant No. 11100265.
\end{acknowledgments}

\end{document}